\documentclass[twocolumn,english]{revtex4-1}
\usepackage[T1]{fontenc}
\usepackage[latin9]{inputenc}
\usepackage{amsmath}
\usepackage{amsthm}
\usepackage{amssymb}
\usepackage{graphicx}
\usepackage{color}

\makeatletter
\theoremstyle{plain}
\newtheorem{prop}{\protect\propositionname}

\usepackage{braket}
\usepackage{amsmath}

\makeatother

\usepackage{babel}
\providecommand{\propositionname}{Proposition}

\begin{document}
\title{Tight bounds from multiple-observable entropic uncertainty relations}

\author{Alberto Riccardi}
\affiliation{INFN Sezione di Pavia, Via Agostino Bassi 6, I-27100 Pavia, Italy}

\author{Giovanni Chesi}
\affiliation{INFN Sezione di Pavia, Via Agostino Bassi 6, I-27100 Pavia, Italy}

\author{Chiara Macchiavello}
\affiliation{Dipartimento di Fisica, Universit\`{a} degli Studi di Pavia, Via Agostino Bassi 6, I-27100, Pavia, Italy \\
INFN Sezione di Pavia, Via Agostino Bassi 6, I-27100, Pavia, Italy}

\author{Lorenzo Maccone}
\affiliation{Dipartimento di Fisica, Universit\`{a} degli Studi di Pavia, Via Agostino Bassi 6, I-27100, Pavia, Italy \\
INFN Sezione di Pavia, Via Agostino Bassi 6, I-27100, Pavia, Italy}

\begin{abstract}

We investigate the additivity properties for both bipartite and multipartite systems by using entropic uncertainty relations (EUR) defined in terms of the joint Shannon entropy of probabilities of local measurement outcomes. In particular, we introduce state-independent and state-dependent entropic inequalities. Interestingly, the violation of these inequalities is strictly connected with the presence of quantum correlations. We show that the additivity of EUR holds only for EUR that involve two observables, while this is not the case for inequalities that consider more than two observables or the addition of the von Neumann entropy of a subsystem. We apply them to bipartite systems and to several classes of states of a three-qubit system.
\end{abstract}

\maketitle

Entropic uncertainty relations (EUR) are inequalities that express
preparation uncertainty relations (UR) as sums of Shannon entropies
of probability distributions of measurement outcomes. First introduced
for continuous variables systems \cite{EUR1,EUR2,EUR3,EUR4}, they
were then generalized for pair of observables with discrete spectra
\cite{EUR5,EUR6,EUR7,EUR9,EUR10} (see \cite{EUR8} for a review of
the topic). Conversely to the most known URs defined for product of
variances \cite{Heis1,Robertson1}, which are usually state-dependent,
EURs provide lower bounds, which quantify the knowledge trade-off between
the different observables, that are state-independent. 
Variance-based URs for the sum of variances \cite{SVar4} in some cases
also provide state-independent bounds \cite{SVar1,SVar2,SVar3,SVar5,SVar6}, but
EURs, due to their simple structure, allow to consider URs for more
than two observables in a natural way by simply adding more entropies,
a task that is not straightforward for URs based on the product of variances. However,
tight bounds for multiple-observable EURs are known only for small
dimensions and for restricted sets of observables, typically for complementary
observables \cite{MultiAzarchs,MultiBallesterWehner,MultiIvanovic,MultiSanchez,TightAR,MultiCHina,MultiMolner},
namely the ones that have mutually unbiased bases as eigenbases, and
for angular momentum observables \cite{URAngular-1,TightAR}. \\
Besides their importance from a fundamental point of view as preparation
uncertainty relations, EURs have recently been used to investigate
the nature of correlations in composite quantum systems, providing
criteria that enable to detect the presence of different types of
quantum correlations, both for bipartite and multipartite systems.
Entanglement criteria based on EURs were defined in \cite{EntGuh,Ent1,Ent2,Ent3,Huang},
while steering inequalities in \cite{Steering1,Steering2,Steering3,Steering4,Steering5,SteeringAR}.
Almost all of these criteria are based on EURs for conditional Shannon
entropies, where one tries to exploit, in the presence of correlations,
side information about some subsystems to reduce global uncertainties,
while only partial results for joint Shannon entropies are known \cite{EntGuh,Ent4}.
Moreover, it has been recently proven in \cite{Additivity} that if
one considers EURs defined for the joint Shannon entropy and only pairs
of observables, then it is not possible to distinguish between separable and
entangled states since in this case additivity holds, namely, the lower bound on the sum of the joint Shannon entropies $H(A_1,B_1)$ and $H(A_2,B_2)$ is given by the sum of the lower bounds on the sum of the entropies $H(X_1)$ and $H(X_2)$, with $X = A,B$. \\
In this paper we show that if we consider EURs for more than two observables
the additivity of EURs does no longer hold. This result implies that it is possible
to define criteria that certify the presence of entanglement by using
the joint Shannon entropy for both the bipartite and the multipartite
case. We investigate which criteria can be derived from EURs based on the joint Shannon entropy and their performance. We then provide some examples of entangled states that violate our criteria.
This paper is organized as follows: in Section I we briefly review
some concepts of single system EURs, in particular we discuss the case
of multiple observables. In Section II we establish the entanglement
criteria for bipartite systems and in Section III we address the problem
in the multipartite scenario. In the Appendix, we consider some
examples of entangled states that are detected by these criteria, in
particular we focus on the multi-qubit case. 

\section{Entropic uncertainty relations: a brief review}

The paradigmatic example of EUR for observables with a discrete
non-degenerate spectrum is due to Maassen and Uffink \cite{EUR7},
and it states that for any two observables $A_{1}$ and $A_{2}$,
defined on a $d$-dimensional system, the following inequality holds:
\begin{equation}
H(A_{1})+H(A_{2})\geqslant-2\log_{2}c=q_{MU},\label{MUF}
\end{equation}
where $H(A_{1})$ and $H(A_{2})$ are the Shannon entropies of the
measurement outcomes of two observables $A_{1}=\sum_{j}a_{j}^{1}\ket{a_{j}^{1}}\bra{a_{j}^{1}}$
and $A_{2}=\sum_{j}a_{j}^{2}\ket{a_{j}^{2}}\bra{a_{j}^{2}}$, namely
$H(A_{I})=-\sum_{j}p(a_{j}^{I})\log p(a_{j}^{I})$ being $p(a_{j}^{I})$
the probability of obtaining the outcome $a_{J}^{I}$ of $A_{I}$,
and $c=\max_{j,k}\left|\braket{a_{j}^{1}|a_{k}^{2}}\right|$ is the
maximum overlap between their eigenstates. The bound in Eq.~(\ref{MUF})
is known to be tight if $A_{1}$ and $A_{2}$ are complementary observables \cite{EUR8}.
We remind that two observables $A_{1}$ and $A_{2}$ are said to be
complementary iff their eigenbases are mutually unbiased, namely iff
$\left|\braket{a_{j}^{1}|a_{k}^{2}}\right|=\frac{1}{\sqrt{d}}$ for
all eigenstates, where $d$ is the dimension of the system (see \cite{MUBs}
for a review on MUBs). In this case $q_{MU}=\log_{2}d$, hence we
have:
\begin{equation}
H(A_{1})+H(A_{2})\geqslant\log_{2}d.
\end{equation}
The above relation has a clear interpretation as UR:
let us suppose that $H(A_{1})=0$, which means that the state of the
system is an eigenstate of $A_{1}$, then the other entropy $H(A_{2})$
must be maximal, hence if we have a perfect knowledge of one observable
the other must be completely undetermined. For arbitrary observables
stronger bounds, that involve the second largest term in $\left|\braket{a_{j}|b_{k}}\right|$,
were derived in \cite{EUR10,EUR9}. 

An interesting feature of EURs is that they can be generalized
to an arbitrary number of observables in a straightforward way from
Maassen and Uffink's EUR. Indeed, let us consider for simplicity the
case of three observables $A_{1}$, $A_{2}$ and $A_{3}$, which
mutually satisfy the following EURs:
\begin{equation}
H(A_{i})+H(A_{j})\geqslant q_{MU}^{ij},\label{MUF2}
\end{equation}
where $i,j=1,2,3$ labels the three observables. Then, we have:
\begin{align}
\sum_{k=1}^{3}H(A_{k}) & =\frac{1}{2}\sum_{k=1}^{2}\sum_{j = k+1}^3 \left[H(A_{k})+H(A_{j})\right]\nonumber \\
 & \geq\frac{1}{2}\left(q_{MU}^{12}+q_{MU}^{13}+q_{MU}^{23}\right)
\end{align}
where we have applied (\ref{MUF2}) to each pair. If we have $L$
observables, the above inequality becomes: 
\begin{equation}
\sum_{k=1}^{L}H(A_{k})\geq\frac{1}{\left(L-1\right)}\sum_{t\in T_{2}}q_{MU}^{t},\label{MultiObsEUR}
\end{equation}
where $t$ takes values in the set $T_{2}$ of labels of all the
possible $L(L-1)/2$ pairs of observables. For example if
$L=4$, then $T_{2}=\{12,13,14,23,24,34\}$. However, EURs in the form
(\ref{MultiObsEUR}) are usually not tight, i.e. in most cases the
lower bounds can be improved. Tight bounds are known only for small
dimensions and for complementary or angular momentum observables.
For the sake of simplicity, henceforth all explicit examples will
be discussed only for complementary observables. The maximal number
of complementary observables for any given dimension is an open problem
\cite{MUBs}, which finds its roots in the classification of all complex
Hadamard matrices. However, if $d$ is a power of a prime then $d+1$
complementary observables always exist. For any $d$, even
if it is not a power of a prime, it is possible to find at least three
complementary observables \cite{MUBs}. The method that we will define
in the next Section can be therefore used in any dimension. The qubit
case, where at most three complementary observables exist, which are
in correspondence with the three Pauli matrices, was studied in \cite{MultiSanchez},
while for systems with dimension three to five tight bounds for an
arbitrary number of complementary observables were derived in \cite{TightAR}.
For example in the qubit case, where the three observables $A_{1},A_{2}$
and $A_{3}$ correspond to the three Pauli matrices $\sigma_{x},\sigma_{y}$
and $\sigma_{z},$ we have:
\begin{equation}
H(A_{1})+H(A_{2})+H\left(A_{3}\right)\geqslant2,\label{MultiQubit}
\end{equation}
and the minimum is achieved by the eigenstates of one of the $A_{i}$.
In the case of a qutrit, where four complementary observables exist,
we instead have:
\begin{align}
 & H(A_{1})+H(A_{2})+H(A_{3})\geqslant3,\label{3d 3Mubs}\\
 & H(A_{1})+H(A_{2})+H(A_{3})+H(A_{4})\geqslant4.\label{3d 4mubs}
\end{align}
The minimum values are achieved by:
\begin{align}
 & \frac{e^{i\varphi}\ket0+\ket1}{\sqrt{2}},\ \frac{e^{i\varphi}\ket0+\ket2}{\sqrt{2}},\ \frac{e^{i\varphi}\ket1+\ket2}{\sqrt{2}},
\end{align}
where $\varphi=\frac{\pi}{3},\pi,\frac{5\pi}{3}$. Another result,
for $L<d+1$, can be found in \cite{MultiBallesterWehner}, where
it has been shown that if the Hilbert space dimension is a square,
that is $d=r^{2},$ then for $L<r+1$ the inequality (\ref{MultiObsEUR})
is tight, namely: 
\begin{equation}
\sum_{i=1}^{L}H(A_{i})\geqslant\frac{L}{2}\log_{2}d=q_{BW}.\label{Ballester}
\end{equation}
In order to have a compact expression to use, we express the EUR for
$L$ observables in the following way:
\begin{equation}
\sum_{i=1}^{L}H(A_{i})\geq f\left(\mathcal{A},L\right),\label{L-ObsEur}
\end{equation}
where $f\left(\mathcal{A},L\right)$ indicates the lower bound, which
can be tight or not, and it depends on the set $\mathcal{A}=\left\{ A_{1},...,A_{L}\right\} $
of $L$ observables considered. Here we also point out in the lower
bound how many observables are involved. When we refer explicitly to
tight bounds we will use the additional label $T$, namely $f^{T}\left(\mathcal{A},L\right)$
expresses a lower bound that we know is achievable via some states. 

\section{Bipartite entanglement criteria}

In this Section, we discuss bipartite entanglement criteria based on
EURs, defined in terms of joint Shannon entropies. The framework consists
in two parties, say Alice and Bob, who share a quantum state $\rho_{AB}$,
and they want to establish if their state is entangled. Alice
and Bob can perform $L$ measurements each, that we indicate respectively
as $A_{1},..,A_{L}$ and $B_{1},..,B_{L}$. Alice and Bob measure the
observables $A_{i}\otimes B_{j}$ and they want to have a criterion
defined in terms of the joint Shannon entropies $H\left(A_{i},B_{j}\right)$
which certifies the presence of entanglement. As a reminder, in a
bipartite scenario we say that the state $\rho_{AB}$ is entangled
iff it cannot be expressed as a convex combination of product states,
which are represented by separable states, namely iff: 
\begin{equation}
\rho_{AB}\neq\sum_{i}p_{i}\rho_{A}^{i}\otimes\rho_{B}^{i},
\end{equation}
where $p_{i}\geq0$, $\sum_{i}p_{i}=1$, and $\rho_{A}^{i}$, $\rho_{B}^{i}$
are Alice and Bob's states respectively. 
\begin{prop}
If the state $\rho_{AB}$ is separable, then the following EUR must
hold:
\begin{equation}
\sum_{i=1}^{L}H(A_{i},B_{i})\geq f\left(\mathcal{A},L\right)+f\left(\mathcal{B},L\right),\label{Primo criterio}
\end{equation}
where $f\left(\mathcal{A},L\right)$ and $f\left(\mathcal{B},L\right)$
are the lower bounds of the single system EUR, namely
\begin{equation}
\sum_{i=1}^{L}H(A_{i})\geq f\left(\mathcal{A},L\right),
\end{equation}
\begin{equation}
\sum_{i=1}^{L}H(B_{i})\geq f\left(\mathcal{B},L\right).
\end{equation}
\end{prop}
\begin{proof}
Let us focus first on $H(A_{i},B_{i})$ which, for the properties
of the Shannon entropy, can be expressed as: 
\begin{equation}
H(A_{i},B_{i})=H(A_{i})+H(B_{i}|A_{i}).
\end{equation}
We want to bound $H\left(B_{i}|A_{i}\right)$ which is computed over
the state $\rho_{AB}=\sum_{j}p_{j}\rho_{A}^{j}\otimes\rho_{B}^{j}$
. Through the convexity of the relative entropy, one can prove that
the conditional entropy $H(B|A)$ is concave in $\rho_{AB}$. Then we
have:

\begin{equation}
H(B_{i}|A_{i})_{\sum_{j}p_{j}\rho_{A}^{j}\otimes\rho_{B}^{j}}\geq\sum_{j}p_{j}H(B_{i}|A_{i})_{\rho_{A}^{j}\otimes\rho_{B}^{j}},
\end{equation}
thus, since the right-hand side of the above Equation is evaluated on a product
state, we have: 

\begin{equation}
H(B_{i}|A_{i})_{\sum_{j}p_{j}\rho_{A}^{j}\otimes\rho_{B}^{j}}\geq\sum_{j}p_{j}H(B_{i})_{\rho_{B}^{j}}.
\end{equation}
Therefore, considering $\sum_{i=1}^{L}H(A_{i},B_{i})$, we derive
the following:
\begin{equation}
\sum_{i=1}^{L}H(A_{i},B_{i})\geq\sum_{i}H(A_{i})+\sum_{j}p_{j}\sum_{i}H(B_{i})_{\rho_{B}^{j}}.\label{Proof}
\end{equation}
Then we can observe that $\sum_{i=1}^{L}H(A_{i})\geq f\left(\mathcal{A},L\right)$
and $\sum_{i}H(B_{i})_{\rho_{B}^{j}}\geq f\left(\mathcal{B},L\right)$,
the latter holding due to EUR being state-independent
bounds. Therefore we have:

\begin{align}
\sum_{i=1}^{L}H(A_{i},B_{i}) & \geq f\left(\mathcal{A},L\right)+\sum_{j}p_{j}f\left(\mathcal{B},L\right)\nonumber \\
 & =f\left(\mathcal{A},L\right)+f\left(\mathcal{B},L\right),
\end{align}
since $\sum_{j}p_{j}=1.$ 
\end{proof}
Any state that violates the inequality $\sum_{i=1}^{L}H(A_{i},B_{i})\geq f\left(\mathcal{A},L\right)+f\left(\mathcal{B},L\right)$
must be therefore entangled. If we consider the observables
$A_{i}\otimes B_{i}$ as ones of the bipartite system then they must
satisfy an EUR for all states, even the entangled ones, which can
be expressed as:
\begin{equation}
\sum_{i=1}^{L}H(A_{i},B_{i})\geq f(\mathcal{AB},L),
\end{equation}
where the lower bound now depends on the observables $A_{i}\otimes B_{i}$,
while $f\left(\mathcal{A},L\right)$ and $f\left(\mathcal{B},L\right)$
depend respectively on $A_{i}$ and $B_{i}$ individually. In order
to have a proper entanglement criterion then we should have that 
\begin{equation}
f(\mathcal{AB},L)<f\left(\mathcal{A},L\right)+f\left(\mathcal{B},L\right),
\end{equation}
which means that the set of entangled states that violate the inequality
is not empty. In the case $L=2$, this is not sufficient to have
a proper entanglement criterion. Indeed, as it was shown in \cite{Additivity}, for $L=2$, we
have $f(\mathcal{AB},2)=f\left(\mathcal{A},2\right)+f\left(\mathcal{B},2\right)$
for any observables, which expresses the additivity of EURs for pairs
of observables. A counterexample of this additivity property for $L>3$
is provided by the complete set of complementary observables for two
qubits, indeed we have:
\begin{equation}
H(A_{1},B_{1})+H(A_{2},B_{2})+H(A_{3},B_{3})\geq3,
\end{equation}
and the minimum is attained by the Bell states while $f\left(\mathcal{A},3\right)+f\left(\mathcal{B},3\right)=4,$
which provides the threshold that enables entanglement detection
in the case of two qubits. \\
Let us now clarify the difference of this result with respect to those
defined in terms of EURs based on conditional entropies, in particular
to entropic steering inequalities. Indeed, if one looks at the proof
of Proposition 1, it could be claimed that there is no difference at all
since we used the fact that $\sum_{i}H(B_{i}|A_{i})\geq f(\mathcal{B},L)$,
which is a steering inequality, namely a violation of it witnesses the
presence of quantum steering from Alice to Bob. However, the difference
is due to the symmetric behavior of the joint entropy, which contrasts
with the asymmetry of quantum steering. To be more formal, the joint
Shannon entropy $H(A_{i},B_{i})$ can be rewritten in two forms: 
\begin{align}
H(A_{i},B_{i})= & H(A_{i})+H(B_{i}|A_{i})\\
= & H(B_{i})+H(A_{i}|B_{i}),\nonumber 
\end{align}
then:
\begin{equation}
\sum_{i}H(A_{i},B_{i})=\sum_{i}\left[H(A_{i})+H(B_{i}|A_{i})\right],\label{SI 1}
\end{equation}
and 
\begin{equation}
\sum_{i}H(A_{i},B_{i})=\sum_{i}\left[H(B_{i})+H(A_{i}|B_{i})\right].\label{SI2}
\end{equation}
If now the state is not steerable from Alice to Bob, we have $\sum_{i}H(B_{i}|A_{i})\geq f(\mathcal{B},L)$,
which implies $\sum_{i=1}^{L}H(A_{i},B_{i})\geq f\left(\mathcal{A},L\right)+f\left(\mathcal{B},L\right)$.
Note that in this case if we look at $\sum_{i}H(A_{i}|B_{i})$ no
bound can be derived, apart from the trivial bound $\sum_{i}H(A_{i}|B_{i})\geq0$,
since there are no assumptions on the conditioning from Bob to Alice.
Conversely, if the state is not steerable from Bob to Alice, i.e.
we exchange the roles, we have $\sum_{i}H(B_{i}|A_{i})\geq0$ and
$\sum_{i}H(A_{i}|B_{i})\geq f(\mathcal{A},L)$, which implies again
$\sum_{i=1}^{L}H(A_{i},B_{i})\geq f\left(\mathcal{A},L\right)+f\left(\mathcal{B},L\right)$.
Therefore, if we just look at the inequality in Eq.~(\ref{Primo criterio}),
we cannot distinguish between entanglement and the two possible forms
of quantum steering, but, since the presence of steering, for bipartite
systems, implies entanglement, it is more natural to think about Eq.~(\ref{Primo criterio})
as an entanglement criterion, while if we want to investigate steering
properties of the state we should look at the violation of the criteria
$\sum_{i}H(B_{i}|A_{i})\geq f(\mathcal{B},L)$ and $\sum_{i}H(A_{i}|B_{i})\geq f(\mathcal{A},L).$ In other words, Proposition 1 only detects two-way steerable states.

\subsubsection*{State-dependent bounds}

A stronger entanglement criteria can be derived by considering the
state-dependent EUR:
\begin{equation}
\sum_{i=1}^{L}H(A_{i})\geq f\left(\mathcal{A},L\right)+S\left(\rho_{A}\right),\label{State dependent}
\end{equation}
or the corresponding version for Bob's system $\sum_{i=1}^{L}H(B_{i})\geq f\left(\mathcal{B},L\right)+S\left(\rho_{B}\right)$,
where $S\left(\rho_{A}\right)$ and $S\left(\rho_{B}\right)$ are
the Von Neumann entropies of the marginal states of $\rho_{AB}.$ 
\begin{prop}
If the state $\rho_{AB}$ is separable, then the following EUR must
hold:
\begin{equation}
\sum_{i=1}^{L}H(A_{i},B_{i})\geq f\left(\mathcal{A},L\right)+f\left(\mathcal{B},L\right)+\max\left(S\left(\rho_{A}\right),S\left(\rho_{B}\right)\right).\label{Primo criterio-1}
\end{equation}
\end{prop}
\begin{proof}
The proof is the same of Proposition 1 where we use (\ref{State dependent})
in (\ref{Proof}), instead of the state-dependent bound (\ref{L-ObsEur}).
The same holds if we use the analogous version for Bob. Then, aiming
at the strongest criterion, we can take the maximum between the two
Von Neumann entropies.
\end{proof}
The edge in using these criteria, instead of the one defined in Proposition
1, is such that even for $L=2$ the bound is meaningful. Indeed
a necessary condition to the definition of a proper criterion is that:
\begin{equation}
f^{T}(\mathcal{AB},2)<f\left(\mathcal{A},2\right)+f\left(\mathcal{B},2\right)+S\left(\rho_{X}\right),\label{VN criteria}
\end{equation}
where $X=A,B$ with the additional requirement that the bound on the
left is tight, i.e. there exist states the violate the criterion.
As an example, we can consider a two-qubit system, the observables
$X_{AB}=\sigma_{X}^{A}\otimes\sigma_{X}^{B}$ and $Z_{AB}=\sigma_{Z}^{A}\otimes\sigma_{Z}^{B}$,
which for all states of the whole system satisfy $H(X_{AB})+H(Z_{AB})\geq2$,
and the Bell state $\rho_{AB}=\ket{\phi^{+}}\bra{\phi^{+}}$. In
this scenario, the entanglement criterion reads:
\begin{equation}
H(X_{AB})+H(Z_{AB})\geq3,
\end{equation}
which is actually violated since the left-hand side is equal to 2.
Note that in general the condition $f^{T}(\mathcal{AB},L)<f\left(\mathcal{A},L\right)+f\left(\mathcal{B},L\right)+S\left(\rho_{X}\right)$
is necessary to the usefulness of the corresponding entanglement criteria. 

\section{Multipartite entanglement criteria}

We now extend the results of Propositions 1 and 2 for multipartite
systems, where the notion of entanglement has to be briefly discussed
since it has a much richer structure than the bipartite case.
Indeed, we can distinguish among different levels of separability.
First, we say that a state $\rho_{V_{1},..,V_{n}}$ of $n$ systems
$V_{1},..,V_{n}$ is fully separable iff it can be written
in the form:
\begin{equation}
\rho_{V_{1},..,V_{n}}^{FS}=\sum_{i}p_{i}\rho_{V_{1}}^{i}\otimes...\otimes\rho_{V_{n}}^{i},\label{Fully sep}
\end{equation}
with $\sum_{i}p_{i}=1$, namely it is a convex combination of product
states of the single subsystems. As a case of study, we will always
refer to tripartite systems, where there are three parties, say Alice,
Bob and Charlie. In this case a fully separable state can be written
as:
\begin{equation}
\rho_{ABC}^{FS}=\sum_{i}p_{i}\rho_{A}^{i}\otimes\rho_{B}^{i}\otimes\rho_{C}^{i}.
\end{equation}
Any state that does not admit such a decomposition contains entanglement among some subsystems. However, we can
define different levels of separability. Hence, we say that the state $\rho_{V_1,..,V_{n}}$
of $n$ systems is separable with respect to a given partition $\{I_{1},..,I_{k}\}$,
where $I_{i}$ are disjoint subsets of the indices $I=\{1,..,n\}$,
such that $\cup_{j=1}^{k}I_{j}=I$, iff it can be expressed as:
\begin{equation}
\rho_{V_{1},..,V_{n}}^{1,..,k}=\sum_{i}p_{i}\rho_{1}^{i}\otimes..\otimes\rho_{k}^{i},
\end{equation}
where $\rho^i_{\alpha}$ is the state of the system $\{V_i:i\in I_{\alpha}\}$, $\alpha = 1, ..., k$.
Namely, some systems share entangled states, while the state is separable
with respect to the partition considered. For tripartite system we
have three different possible bipartitions: $1|23$, $2|13$ and $3|12$.
As an example, if the state $\rho_{ABC}$ can be expressed as:
\begin{equation}
\rho_{ABC}^{1|23}=\sum_{i}p_{i}\rho_{A}^{i}\otimes\rho_{BC}^{i},
\end{equation}
then there is no entanglement between Alice and Bob+Charlie, while
these last two share entanglement. If a state does not admit such
a decomposition, it is entangled with respect to this partition.
Finally, we say that $\rho_{V_{1},..,V_{n}}$ of $n$ systems can
have at most $m$-system entanglement iff it is a mixture of all states
such that each of them is separable with respect to some partition
$\{I_{1},..,I_{k}\}$, where all sets of indices $I_{k}$ have cardinality
$N\leq m$. For tripartite systems this corresponds to the notion of
biseparability, namely the state can have at most 2-system entanglement.
A biseparable state can be written as:
\begin{equation}
\rho_{ABC}=\sum_{i}p_{i}\rho_{A}^{i}\otimes\rho_{BC}^{i}+\sum_{j}q_{j}\rho_{B}^{j}\otimes\rho_{AC}^{j}+\sum_{k}m_{k}\rho_{C}^{k}\otimes\rho_{AB}^{k},
\end{equation}
 with $\sum_{i}p_{i}+\sum_{j}q_{j}+\sum_{k}m_{k}=1.$ For $n=3$ a state is then said
to be genuine tripartite entangled if it is $3$-system entangled,
namely if it does not admit such a decomposition. 

\subsubsection*{Full separability}

Let us clarify the scenario: In each system $V_{i}$ we consider a
set of $L$ observables $V_{i}^{1},..,V_{i}^{L}$ that we indicate
as $\mathcal{V}_{i}.$ The single-system EUR is expressed as: 
\begin{equation}
\sum_{j=1}^{L}H\left(V_{i}^{j}\right)\geq f\left(\mathcal{V}_{i},L\right).\label{EUR Vi}
\end{equation}
We are interested in defining criteria in terms of $\sum_{j=1}^{L}H\left(V_{1}^{j},..,V_{n}^{j}\right)$.
A first result regards the notion of full separability. 
\begin{prop}
If the state $\rho_{V_{1},..,V_{n}}$ is fully separable, then the
following EUR must hold:
\begin{equation}
\sum_{j=1}^{L}H\left(V_{1}^{j},..,V_{n}^{j}\right)\geq\sum_{i=1}^{n}f\left(\mathcal{V}_{i},L\right).
\end{equation}
\end{prop}
\begin{proof}
Let us consider the case $n=3$. For a given $j$ we have:
\begin{equation}
H\left(V_{1}^{j},V_{2}^{j},V_{3}^{j}\right)=H\left(V_{1}^{j}\right)+H\left(V_{2}^{j}V_{3}^{j}|V_{1}^{j}\right).
\end{equation}
Since the state is separable with respect to the partition $23|1$, due
to concavity of the Shannon entropy, we have:
\begin{equation}
H\left(V_{2}^{j}V_{3}^{j}|V_{1}^{j}\right)\geq\sum_{i}p_{i}H(V_{2}^{j}V_{3}^{j})_{\rho_{2}^{i}\otimes\rho_{3}^{i}}.
\end{equation}
By using the chain rule of the Shannon entropy, the above right-hand side
can be rewritten as:
\begin{align}
\sum_{i}p_{i}H(V_{2}^{j}V_{3}^{j})_{\rho_{2}^{i}\otimes\rho_{3}^{i}}= & \sum_{i}p_{i}H(V_{2}^{j})_{\rho_{2}^{i}}\nonumber \\
 & +\sum_{i}p_{i}H(V_{3}^{j}|V_{2}^{j})_{\rho_{2}^{i}\otimes\rho_{3}^{i}},
\end{align}
where the last term can be lower bounded by exploiting the separability
of the state and the concavity of the Shannon entropy, namely:
\begin{equation}
\sum_{i}p_{i}H(V_{3}^{j}|V_{2}^{j})_{\rho_{2}^{i}\otimes\rho_{3}^{i}}\geq\sum_{i}p_{i}H(V_{3}^{j})_{\rho_{3}^{i}}.
\end{equation}
By summing over $j$ we arrive at the thesis:
\begin{equation}
\sum_{j=1}^{L}H\left(V_{1}^{j},V_{2}^{j},V_{3}^{j}\right)\geq\sum_{i=1}^{3}f\left(\mathcal{V}_{i},L\right),
\end{equation}
since $\sum_{j}H\left(V_{1}^{j}\right)\geq f\left(\mathcal{V}_{1},L\right)$,
$\sum_{i}p_{i}\sum_{j}H(V_{2}^{j})_{\rho_{2}^{i}}\geq f\left(\mathcal{V}_{2},L\right)$
and $\sum_{i}p_{i}\sum_{j}H(V_{3}^{j})_{\rho_{3}^{i}}\geq f\left(\mathcal{V}_{3},L\right)$
because of the state-independent EUR. The extension of the proof to
$n$ systems is straightforward.
\end{proof}
The following proposition follows directly by considering the state-dependent
bound:
\begin{equation}
\sum_{j=1}^{L}H\left(V_{i}^{j}\right)\geq f\left(\mathcal{V}_{i},L\right)+S\left(\rho_{i}\right).\label{dh}
\end{equation}

\begin{prop}
If the state $\rho_{V_{1},..,V_{n}}$ is fully separable, then the
following EUR must hold:
\begin{equation}
\sum_{j=1}^{L}H\left(V_{1}^{j},..,V_{n}^{j}\right)\geq\sum_{i=1}^{n}f\left(\mathcal{V}_{i},L\right)+\max\left(S\left(\rho_{1}\right),...,S\left(\rho_{n}\right)\right).
\end{equation}
\end{prop}
Note that only the von Neumann entropy of one system is present in
the above inequality. This is due to the fact that we use only Eq.~(\ref{dh})
in the first step of the proof, otherwise we would end with criteria
that require the knowledge of the decomposition in Eq.~(\ref{Fully sep}). 

\subsubsection*{Genuine multipartite entanglement}

We now analyze the strongest form of multipartite entanglement in
the case of three systems, say Alice, Bob and Charlie. We make the
further assumptions that the three systems have the same dimension
and in each system the parties perform the same set of measurements,
which implies that there is only one bound $F_{1}\left(L\right)$ given by the single-system EURs
$\sum_{j=1}^{L}H\left(V_{j}\right)\geq F_{1}\left(L\right)$, with $V=A,B,C$. Similarly, we indicate
the bound on a pair of systems as $F_{2}\left(L\right)$, namely $\sum_{j=1}^{L}H\left(V_{j},W_{j}\right)\geq F_{2}\left(L\right)$,
with $V \neq W = A,B,C$. Then, the criterion
defined in Proposition 3 for three systems reads as $\sum_{j=1}^{L}H\left(A_j, B_j, C_j\right)\geq3F_{1}\left(L\right)$,
and must be satisfied by all fully separable states. 
\begin{prop} \label{p5}
If $\rho_{ABC}$ is not genuine multipartite entangled, namely
it is biseparable, then the following EUR must hold:
\begin{equation}
\sum_{j=1}^{L}H\left(A_j, B_j, C_j\right)\geq\frac{5}{3}F_{1}\left(L\right)+\frac{1}{3}F_{2}(L).\label{Prop 5}
\end{equation}
\end{prop}
\begin{proof}
Let us assume that $\rho_{ABC}$ is biseparable, that is:
\begin{equation}
\rho_{ABC}=\sum_{i}p_{i}\rho_{A}^{i}\otimes\rho_{BC}^{i}+\sum_{l}q_{l}\rho_{B}^{l}\otimes\rho_{AC}^{l}+\sum_{k}m_{k}\rho_{C}^{k}\otimes\rho_{AB}^{k}
\end{equation}
with $\sum_ip_i + \sum_lq_l + \sum_km_k = 1$.
The joint Shannon entropy $H\left(A_j, B_j, C_j\right)$
can be expressed as:
\begin{align}
H\left(A_j, B_j, C_j\right)= \, & \frac{1}{3}\left[H(A_j)+H\left(B_j,C_j|A_j\right)\right] + \label{chai rules}\\
 & \frac{1}{3}\left[H(B_j)+H\left(A_j,C_j|B_j\right)\right] + \nonumber \\
 & \frac{1}{3}\left[H(C_j)+H\left(A_j,B_j|C_j\right)\right].\nonumber 
\end{align}
By using the concavity of Shannon entropy and the fact that the state
is biseparable we find the relations
\begin{align}
H\left(B_j,C_j|A_j\right)  \geq & \sum_{i}p_{i}H\left(B_j,C_j\right)_{\rho_{BC}^{i}} +\\
 &\sum_{l}q_{l}H(B_j)_{\rho_{B}^{l}}+\sum_{k}m_{k}H(C_j)_{\rho_{C}^{k}}\nonumber 
\end{align}
\begin{align}
H\left(A_j,B_j|C_j\right) \geq & \sum_{i}p_{i}H\left(A_j\right)_{\rho_{A}^{i}} +\\
 &\sum_{l}q_{l}H(B_j)_{\rho_{B}^{l}}+\sum_{k}m_{k}H(A_j,B_j)_{\rho_{AB}^{k}}\nonumber 
\end{align}
\begin{align}
H\left(A_j,C_j|B_j\right) \geq & \sum_{i}p_{i}H\left(A_j\right)_{\rho_{A}^{i}} +\\
 & \sum_{l}q_{l}H(A_j,C_j)_{\rho_{AC}^{l}}+\sum_{k}m_{k}H(C_j)_{\rho_{C}^{k}}.\nonumber 
\end{align}
Then, by considering the sum over $j$ of the sum of the above entropies,
and using the EURs related to single systems and to pairs of systems, we find:
\begin{equation}
\begin{array}{c}
\sum_{j}H\left(B_{j},C_{j}|A_{j}\right)+H\left(A_{j},B_{j}|C_{j}\right)+H\left(A_{j},C_{j}|B_{j}\right)\\
\geq2F_{1}\left(L\right)+F_{2}(L).
\end{array}
\end{equation}
The thesis (\ref{Prop 5}) is now implied by combining the expression above, Eq.
(\ref{chai rules}) and the following EUR:
\begin{equation}
\sum_{j}H(A_{j})+H(B_{j})+H(C_{j})\geq3F_{1}\left(L\right).\label{EURs}
\end{equation}
\end{proof}
\begin{prop} \label{p6}
If $\rho_{ABC}$ is not genuine multipartite entangled, namely
it is biseparable, then the following EUR must hold:
\begin{equation}
\sum_{j=1}^{L}H\left(A_{j},B_{j},C_{j}\right)\geq\frac{5}{3}F_{1}\left(L\right)+\frac{1}{3}F_{2}(L)+\frac{1}{3}\sum_{x=A,B,C}S\left(\rho_{X}\right).\label{Prop 5-1}
\end{equation}
\end{prop}
The above Proposition follows from the proof of Proposition~\ref{p5}, where we consider the single-system state-dependent EUR.

Note that, in principle, Propositions~\ref{p5} and~\ref{p6} can be extended to account for a larger number of systems and observables by iteratively exploiting the bounds provided by EURs related to smaller numbers of parties, as we did to prove Proposition~\ref{p5}.

\section{Conclusions}

In conclusion, we derived and characterized EURs defined in terms of the joint Shannon entropy, whose violation implies the presence of entanglement. In the case of bipartite systems, we found that EUR entanglement criteria for the joint Shannon entropies require at least three different observables, namely $L > 2$, or, if one considers only two measurements, the addition of the von Neumann entropy of a subsystem, thus showing that the additivity character of the state-independent EURs retrieved in Ref.~\cite{Additivity} holds only for two measurements. We extended our criteria to the case of multipartite systems, which enable us to discriminate between different types of multipartite entanglement. In particular, we established EURs that allow to certify entanglement among an arbitrary number of parties; then, we focused on the case of three systems and derived the EURs that imply the strongest multipartite criteria. In the Appendix, we showed how these criteria perform for both bipartite and multipartite systems, providing several examples of states that are detected by the proposed criteria. 

This material is based upon work supported by the U.S. Department of Energy, Office of Science, National Quantum Information Science Research Centers, Superconducting Quantum Materials and Systems Center (SQMS) under contract number DEAC02-07CH11359 and by the EU H2020 QuantERA ERA-NET Cofund in Quantum Technologies project QuICHE.

\appendix

\section{}

Here we discuss our criteria for bipartite and multipartite systems. We will mainly focus on pure states and multi-qubit systems. We inspect in detail how many entangled states and which levels of separability can be detected with the different criteria derived from the EURs. We point out that, if one focuses just on the entanglement-detection efficiency, bounds retrieved from EURs based on the joint Shannon entropy are not as good as some existing criteria. On the other hand, note that the experimental verification of our EUR-based criteria may require less measurements. For instance, if we want to detect the entanglement of a multipartite state through the PPT method, we need to perform a tomography of the state, which involves the measurement of $d^4$ observables. On the contrary, the evaluation of the entropies just needs the measurements of the observables involved in the EUR and its number can be fixed at $3$ independently of the dimension $d$.

\subsection{Bipartite systems}

\begin{figure}
{\includegraphics[scale=0.5]{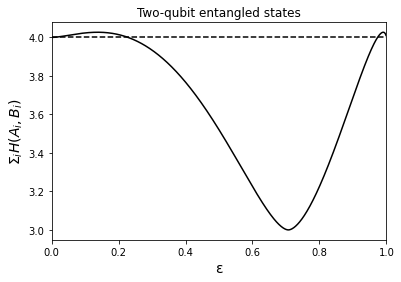}}
\caption{Pure Bipartite States: the continuous line represents $\sum_{i=1}^{3}H(A_{i},B_{i})$ for
the states $\ket{\psi_{\epsilon}}=\epsilon\ket{00}+\sqrt{1-\epsilon^{2}}\ket{11}$
while the dotted line is the bound according to (\ref{criterio2}).
A small set of entangled states is therefore not detected by the criterion
(\ref{criterio2}).}
\label{fig1}
\end{figure}

Let us start with the simple case of two qubits. In this scenario we will consider complementary observables and the single-system tight EURs in Refs. \cite{MultiSanchez,TightAR}.
Hence, in this case the criteria proved in Section II read
\begin{equation}
H(A_{1},B_{1})+H(A_{2},B_{2})<2+\max(S(\rho_{A}),S(\rho_{B})),\label{criterio1}
\end{equation}
\begin{equation}
\sum_{i=1}^{3}H(A_{i},B_{i})<4,\label{criterio2}
\end{equation}

\begin{align}
\sum_{i=1}^{3}H(A_{i},B_{i}) & <4+\max(S(\rho_{A}),S(\rho_{B})),   \label{criterio3}
\end{align}
where $A_{1}=Z_{1}$ ($B_1 = Z_2$), $A_{2}=X_{1}$ ($B_2 = X_2$) and $A_{3}=Y_{1}$ ($B_3 = Y_2$) being $Z_{1}$ ($Z_2$), $X_{1}$ ($X_2$)
and $Y_{1}$ ($Y_2$) the usual Pauli matrices for the first (second) qubit. In the case of two qubits we have already shown in Section II that maximally entangled states are detected by the above criteria.\\
We can then consider the family of entangled two-qubit states given by:
\begin{equation}
\ket{\psi_{\epsilon}}=\epsilon\ket{00}+\sqrt{1-\epsilon^{2}}\ket{11},\label{entangled states1}
\end{equation}
where $\epsilon\in(0,1).$ We first note that for this family we have
$S(\rho_{A})=S(\rho_{B})=-\epsilon^{2}\log_{2}\epsilon^{2}-(1-\epsilon^{2})\log_{2}(1-\epsilon^{2}),$
which is equal to $H(A_{1},B_{1}).$ Conversely, we have instead
$H(A_{2},B_{2})=-\frac{1}{2}(1-\bar{\epsilon})\log_{2}(\frac{1}{4}(1-\bar{\epsilon}))-\frac{1}{2}(1+\bar{\epsilon})\log_{2}(\frac{1}{4}(1+\bar{\epsilon})),$
with $\bar{\epsilon}=2\epsilon\sqrt{1-\epsilon^{2}}$ and $H(A_{2},B_{2})=H(A_{3},B_{3})$.
The family of states in Eq.~(\ref{entangled states1}) is then completely
detected by the criteria in Eqs.~(\ref{criterio1}) and~(\ref{criterio3}), since $H(A_{2},B_{2})<2$, while the inequality in Eq.~(\ref{criterio2}) fails to detect all the states parametrizes by $\varepsilon$ in Eq.~(\ref{entangled states1}), as shown in Fig.~\ref{fig1}.

Let us consider now the entangled two-qudit states given by:
\begin{equation}
\ket{\psi_{\lambda}}=\sum_{i=0}^{d-1}\lambda_{i}\ket{ii},\label{entangled states1-1}
\end{equation}
where $\sum_{i}\lambda_{i}^{2}=1$ and $0<\lambda_{i}<1 \,\,\forall i\in[0,d)$. As an entanglement
criterion we consider:
\begin{equation}
H(A_{1},B_{1})+H(A_{2},B_{2}) < 2\log_{2}d+\max(S(\rho_{A}),S(\rho_{B})),\label{qudit criterion}
\end{equation}
where the observables $A_{1}$ and $B_{1}$ are the computational bases, while
$A_{2}$ and $B_{2}$ the corresponding Fourier transforms. First, we observe that
for these states we have $S(\rho_{A})=S(\rho_{B})=-\sum_{i}\lambda_{i}^{2}\log_{2}\lambda_{i}^{2} = H(A_{1},B_{1})$.
Hence, the entanglement condition becomes:
\begin{equation}
H(A_{2},B_{2})<2\log_{2}d.
\end{equation}

However, for any two-qudit states we have $H(A_{2},B_{2})\leq2\log_{2}d$
and the maximum is achieved by states that give uniform probability
distributions for $A_{2}\otimes B_{2}$. Being $A_{2}$ and $B_{2}$ the Fourier transformes of the computational bases, the family in Eq.~(\ref{entangled states1-1}) cannot give a uniform probability distribution. The maximum value could
be attained only by states of the form $\ket{ii}$, hence by separable
states. Thus, our criterion in Eq.~(\ref{qudit criterion}) detects all two-qudit entangled
states belonging to the family in Eq.~(\ref{entangled states1-1}).\\

\begin{figure}
{\includegraphics[scale=0.33]{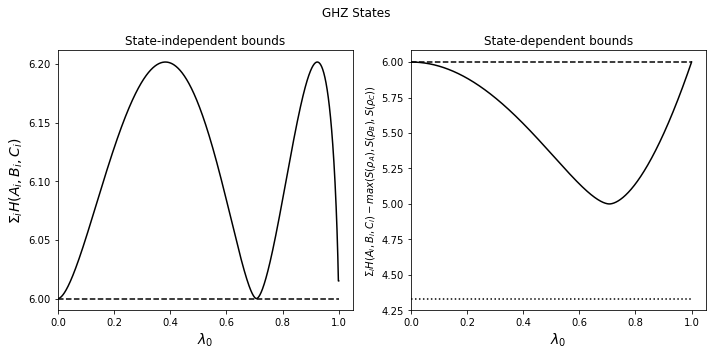}}
\caption{GHZ states. In the left panel, the continuous line represents $\sum_{i=1}^{3}H(A_{i},B_{i},C_{i})$
while the dashed line is the bound according to Eq.~(\ref{multi_ent1}).
Note that this criterion does not identify any entangled state. In the
right panel, the continuous line shows $\sum_{i=1}^{3}H(A_{i},B_{i},C_{i})-\max(S(\rho_{A}),S(\rho_{B}),S(\rho_{C}))$,
while the dashed line represents the bound in Eq.~(\ref{multi_ent2}) and
the lower dotted line the bound in Eq.~(\ref{gen_ent2}), since for
this class of states we have $\max(S(\rho_{A}),S(\rho_{B}),S(\rho_{C}))=\frac{1}{3}\sum_{x=A,B,C}S\left(\rho_{X}\right)$.
By using state-dependent bounds, we correctly detect all non-separable
GHZ states, but we cannot detect which states are genuine
multipartite entangled.}
\label{fig2}
\end{figure}

\subsection{Multipartite systems}

As an example of a multipartite system we focus on the case of three-qubit systems. In this case a straightforward generalization of the
Schmidt decomposition is not available. However, the pure states can be
parameterized and classified in terms of five real parameters: 
\begin{equation}
\ket{\psi}=\lambda_{0}\ket{000}+\lambda_{1}e^{i\phi}\ket{100}+\lambda_{2}\ket{101}+\lambda_{3}\ket{110}+\lambda_{4}\ket{111},
\end{equation}
where $\sum_{i=0}^4\lambda_{i}^{2}=1$. In particular, we are interested
in two classes of entangled states: the GHZ states, given by 
\begin{equation}
\ket{GHZ}=\lambda_{0}\ket{000}+\lambda_{4}\ket{111},
\end{equation}
and the $W$-states, which are
\begin{equation}
\ket{W}=\lambda_{0}\ket{000}+\lambda_{2}\ket{101}+\lambda_{3}\ket{110}.
\end{equation}

\begin{figure}
{\includegraphics[scale=0.5]{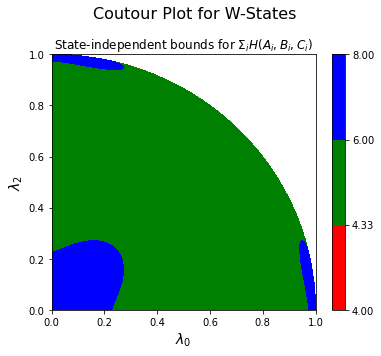}}
\caption{Tripartite W states. The contour plot shows in which areas in the plane $\lambda_{0}\times\lambda_{2}$
the sum of the three entropies $\sum_{i=1}^{3}H(A_{i},B_{i},C_{i})$
is below the state-independent bounds in Eqs.~(\ref{multi_ent1}) and ~(\ref{gen_ent1}). The green area represents
non-separable states, the blue one states that are not identified
by these criteria. No states are identified as genuine multipartite entangled.}
\label{fig3}
\end{figure}

\begin{figure}
{\includegraphics[scale=0.5]{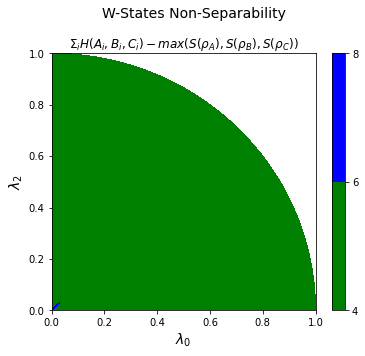}}
\caption{W-state non-separability. The contour plot shows the effectiveness of the state-dependent criterion
(\ref{multi_ent2}) on the W states. Indeed, almost all states are correctly detected (green area) as non-separable. Only a small area (blue) close to the origin is not identified. }
\label{fig4}
\end{figure}

\begin{figure}
{\includegraphics[scale=0.5]{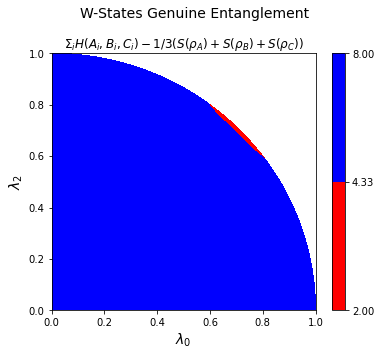}}
\caption{W-State genuine multipartite entanglement. The contour plot shows the performance of the state-dependent criterion in Eq.~(\ref{gen_ent2})
on the W states. A small set of these states is identified as genuine multipartite
entangled (red area).}
\label{fig5}
\end{figure}

The three observables considered in each system
are the Pauli matrices, i.e. $A_{1}=Z_{1},A_{2}=X_{1}$ and
$A_{3}=Y_{1}$ and the same for the other subsystems. The criteria
for detecting the presence of entanglement, namely states that are not fully separable,
in this case are
\begin{equation}
\sum_{i=1}^{3}H(A_{i},B_{i},C_{i})<6 \label{multi_ent1}
\end{equation}
and
\begin{equation}
\sum_{i=1}^{3}H(A_{i},B_{i},C_{i})<6+\max(S(\rho_{A}),S(\rho_{B}),S(\rho_{C})),\label{multi_ent2}
\end{equation}
while the criteria for genuine multipartite entanglement are
\begin{equation}
\sum_{i=1}^{3}H(A_{i},B_{i},C_{i})<\text{\ensuremath{\frac{13}{3}}} \label{gen_ent1}
\end{equation}
and
\begin{align}
\sum_{i=1}^{3}H(A_{i},B_{i},C_{i}) & <\ensuremath{\frac{13}{3}}+\frac{1}{3}\sum_{x=A,B,C}S\left(\rho_{X}\right).\label{gen_ent2}
\end{align}

For the class of GHZ states the sum of the three entropies $\sum_{i=1}^{3}H(A_{i},B_{i},C_{i})$
is plotted as a function of $\lambda_{0}$ in Fig.~\ref{fig2} with respect
to the state-independent and dependent bounds. We can see
that in this case the state-independent bounds fail to detect even the weakest
form of entanglement. Conversely, the state-dependent bounds identify
all states as non-separable but none as genuine multipartite entangled.
\\
For the class of the W states the effectiveness of our criteria is
shown in Figs.~\ref{fig3},~\ref{fig4} and~\ref{fig5}. Since the W states depend on two parameters, here
we use contour plots in the plane $\lambda_{0}\times\lambda_{2}$
showing which subsets of W states are detected as non-fully separable or
genuine multipartite entangled. As we can see, the state-independent
bounds (Fig.~\ref{fig3}) detect the non-separable character for a large subset of W states. Conversely, no state is identified as
genuine multipartite entangled. By using the state-dependent bounds
(Figs.~\ref{fig4} and~\ref{fig5}) we are able to detect almost all non separable W states
and, above all, we can also identify a small subset of W states as
genuine multipartite entangled.

\end{document}